\documentstyle[11pt]{article}

\newcommand{\diff}{d} \newcommand{\lodiff}{s} \begin{document}
\begin{titlepage}

\begin{flushright} ULB--TH--98/03\\ {F}ebruary 1998\\ \end{flushright}

\vfill

\begin{center} {\Large{\bf Local BRST cohomology of the gauged
principal non-linear sigma model }} \end{center}

\vfill

\begin{center} {\large Marc Henneaux\,$^{a,b,1}$\ and Andr\'{e}
 Wilch\,$^{a,2}$} \end{center}

\vfill

\begin{center} {\sl $^a$ {F}acult\'e des Sciences, Universit\'e Libre
 de Bruxelles,\\ Campus Plaine C.P.  231, B--1050 Bruxelles,
 Belgium\\[1.5ex] $^b$ Centro de Estudios Cient\'\i ficos de
 Santiago,\\ Casilla 16443, Santiago 9, Chile } \end{center}

\vfill

\begin{abstract} The local BRST cohomology of the gauged non-linear
sigma model on a group manifold is worked out for any Lie group $G$.
We consider both, the case where the gauge field is dynamical and the
case where it has no kinetic term ($G/G$ topological theory).  Our
results shed a novel light on the problem of gauging the WZW term as
well as on the nature of the topological terms introduced a few years
ago by De Wit, Hull and Ro\v{c}ek.  We also consider the BRST
cohomology of the rigid symmetries of the ungauged model and recover
the results of D'Hoker and Weinberg on the most general effective
actions compatible with the symmetries.  \end{abstract}

\vspace{5em}

{\small \noindent \noindent $^1$ E-mail:  henneaux@ulb.ac.be\\
\noindent $^2$ E-mail:  awilch@ulb.ac.be}

\end{titlepage}


\section{Introduction} \setcounter{equation}{0}

A central theorem in the renormalization of Yang-Mills gauge models
interacting with matter is that the most general solution of the BRST
invariance condition $s \int m \ dx = 0$ is given, up to trivial
terms, by the integral of a gauge invariant polynomial in the field
strengths, the matter fields and their covariant derivatives (in odd
dimensions, there are also Chern-Simons terms).  This theorem
guarantees that all the divergencies appearing in the quantum theory
can be absorbed by counterterms that respect the original symmetry,
making the theory renormalizable in the ``modern sense" in any number
of spacetime dimensions \cite{wein1,wein2} (for related, but different
ideas, see \cite{ansel}).  It also guarantees that gauge invariant
operators can be renormalized in a gauge independent way.

The theorem, conjectured in \cite{klu}, was recently proved in
\cite{hen0,hen1} through cohomological arguments (see \cite{Jogl} for
earlier developments).  Its interest transcends the question of
renormalization since the BRST invariance condition also determines
the allowed deformations of the action, i.e.  the terms that can be
consistently added to the classical, gauge invariant action while
maintaining the number of (possibly deformed) gauge symmetries
\cite{hen2}.

An important assumption made in \cite{hen1} was, however, that the
matter fields transform in linear representations of the symmetry.
Now, non-linear realizations are also important since non-linear sigma
models coupled to gauge fields occur in supergravity, string theory as
well as in the effective description of low energy interactions among
hadrons.

In this paper, we generalize the above theorem to the case of the
gauged non-linear sigma model on a group manifold $G$ (gauged
principal sigma model).  {F}or definiteness, the right action of $G$
on itself will be gauged.  To simplify the discussion, the spacetime
manifold is assumed to be homeomorphic to $R^n$, leaving only the
group manifold as a source for non-trivial topology.  \footnote{{F}or
a discussion on how to take into account the spacetime topology
(restricted to product bundles), see \cite{DV}.}  We show that the
most general deformation of the action is, up to trivial
redefinitions, the integral of a strictly gauge invariant term plus
winding number terms.  (Winding number terms are characterized by two
features:  (i) They involve only group-valued fields, and (ii) they do
not contribute to the field equations but are not exact in field space
and hence cannot be eliminated globally by adding a total derivative.
These terms are related to the De Rham cohomology of the group
manifold - see section \ref{winding}).  In particular, we recover from
a different perspective the fact that there is no room in the
principal case for the gauged Wess-Zumino-Witten term.  {F}urthermore,
we verify explicitly that the Chern-Simons terms actually differ from
strictly gauge-invariant terms by (non-invariant) total derivatives
plus, possibly, winding number terms, even when the Lie algebra
cohomology of ${\cal G}$ is non-trivial (${\cal G}$ denotes the Lie
algebra of $G$).  This property also holds for the topological terms
described in \cite{wit}, which are equivalent to winding number terms
plus strictly gauge-invariant terms in the principal case considered
here.

At the quantum level, our result implies that the coupled
Yang-Mills-non-linear-$\sigma$-model in any number of spacetime
dimensions, even though generically not power-counting renormalizable,
is renormalizable in the \linebreak ``modern sense" of \cite{wein2}.
Note that in perturbation theory, it is customary to restrict the
fields to a neighbourhood of the identity, so that the winding number
terms, which are locally trivial, may be dropped.

We also compute the BRST cohomology for other values of the ghost
degree.  This is relevant for the problem of anomalies in Yang-Mills
theory since the Wess-Zumino compensating field precisely transforms
non-linearly as a group element under gauge transformations
\cite{wess,zumi}.  {F}inally, the cohomology of the ungauged model is
analysed, which enables us to recover from a different angle the
results of D'Hoker and Weinberg on the most general effective actions
compatible with the rigid $G$-symmetry of the $\sigma$-model
\cite{wein3}.

\section{The model} \setcounter{equation}{0}

The starting point of our analysis is a general action of the form
\begin{equation} S_0[A^a_{\mu},g,y^i] = \int {\cal L}(A^a_{\mu},g,y^i)
\ dx , \end{equation} where $A^a_{\mu}$ denotes a Yang-Mills
connection ($a=1, \dots, N$) and $g$ is an element of the
corresponding Lie group $G$.  We assume that $g$ belongs to some
faithful $k \times k$ matrix representation of $G$ and adopt matrix
notations throughout for $g$.  Unless $G = GL(k)$, the matrix elements
of $g$ are not independent.  One may express them in terms of local
coordinates $h^a$ on the group, $g = g(h^a)$, but because this can
usually not be done globally, we shall avoid explicit
parametrizations.  The $y^i$ stand for matter fields that transform
linearly under some representation of $G$ with generators $(Y_a)^i_j$.
We shall also often adopt matrix notations for $Y_a$, viewing the
$y$'s as column-vectors.

At this stage, we do not specify the exact form of the action but only
assume that it is invariant under the following gauge transformations,
\begin{eqnarray} \delta_{\epsilon} A^a_{\mu} &=&
\partial_{\mu}\epsilon^a+f^a_{\ bc}A^b_{\mu}\epsilon^c,\\
\delta_{\epsilon} g &=& g T_a \epsilon^a,\\ \delta_{\epsilon} y &=&
Y_a y \epsilon^a, \end{eqnarray} and that these transformations form a
complete set of gauge symmetries.  The $T_a$ are the generators of the
Lie algebra of $G$ and $f^a_{\ bc}$ are the corresponding structure
constants.

It is convenient to introduce the flat connection $\Theta^a_\mu$
defined through \begin{equation}
g^{-1}dg=\Theta^a_{\mu}T_adx^{\mu}=\Theta^aT_a, \label{flat}
\end{equation} which in terms of local coordinates $h^a$ on $G$ reads
\begin{equation} \Theta^a=\omega^a_{\ b}(h)dh^b.  \end{equation} The
$\omega^a_{\ b}(h)$ are the components of the left-invariant forms
$\Theta^a$ in the basis of the $dh^a$.  The matrix $\omega^a_{\ b}(h)$
is invertible because the invariant forms $\Theta^a$ form a basis.  We
shall denote its inverse by $\Omega^a_{\ b}(h)$, \begin{equation}
\omega^a_{\ b}(h) \Omega^b_{\ c}(h) = \delta^a_c.  \end{equation} The
invariant forms $\Theta^a$ obey the Maurer-Cartan equation,
\begin{equation} d\Theta^a=-\frac{1}{2}f^a_{\ bc}\Theta^b\Theta^c.
\label{MaurerCartan} \end{equation} It follows that the curvature of
the connection (\ref{flat}) vanishes identically, \begin{equation}
{F}^a_{\mu \nu}(\Theta)=\partial_{[\mu}\Theta^a_{\nu]}+ f^a_{\
bc}\Theta^b_{\mu}\Theta^c_{\nu}=0.  \end{equation} The quantity
$I^a_{\mu}$, \begin{equation} I^a_\mu = \Theta^a_\mu - A^a_\mu,
\label{I} \end{equation} transforms homogeneously since it is the
difference between two connections, \begin{equation}
\delta_{\epsilon}I^a_{\mu}=f^a_{\ bc}I^b_{\mu}\epsilon^c,
\end{equation} and it can be thought of as some sort of covariant
derivative of the field $g$ (the notation $I^a_\mu = \omega^a_{\ b}(h)
D_\mu h^b$ is sometimes used in the literature).  Clearly, all the
first-order derivatives $\partial_{\mu}h^a$ of $h^a$ can be expressed
in terms of $I^a_{\mu}$.  The connection $\Theta^a_{\mu}$ can be used
to define covariant derivatives of fields transforming linearly under
the symmetry; to avoid confusion with the $A$-covariant derivative, we
shall denote the corresponding covariant derivative by
$D^{(\Theta)}_\mu$.

There exist two important choices for the action.  One is the standard
gauged model where both the group-valued field and the Yang-Mills
field (as well as the matter fields $y^i$ if any) have a kinetic term,
\begin{equation} {\cal
L}=-\frac{1}{4}g_{ab}{F}^a_{\mu\nu}{F}^{b\mu\nu}
-\frac{1}{2}g_{ab}I^a_{\mu}I^{b\mu} + \mbox{matter action}.
\label{YMaction} \end{equation} Here, $g_{ab}$ is an invertible,
invariant metric on ${\cal G}$, which we assume to exist.  The field
equations are (dropping the matter part) \begin{eqnarray}
D^{(A)}_{\rho}{F}^{a\rho\mu}+I^{a\mu} &=& 0, \label{YMeqn}\\
D^{(\Theta)}_{\mu}I^{a\mu} &=& 0.  \label{Ieqn} \end{eqnarray} The
first equation follows from varying the vector potential while the
second equation is obtained by varying the group element (if one
varies $h^a$, one really obtains (\ref{Ieqn}) multiplied by the matrix
$\omega^a_{\ b}$, which is invertible).  In (\ref{Ieqn}), the
covariant derivative $D^{(\Theta)}_{\mu}$ may be replaced by
$D^{(A)}_{\mu}$ since the $I$'s commute while the structure constants
$f^a_{\ bc}$ are antisymmetric in $b$ and $c$.  This leads to the
alternative form of the equations of motion (\ref{Ieqn}),
\begin{equation} D^{(A)}_{\mu}I^{a\mu} = 0, \label{Ieqn'}
\end{equation} which clearly exhibits that the $g$-equations of motion
are a consequence of the Yang-Mills equations of motion ((\ref{YMeqn})
implies (\ref{Ieqn'}) by taking the covariant divergence with
$D^{(A)}_{\mu}$).  Note the interesting feature that the Yang-Mills
equations alone are independent even though the combined system
(\ref{YMeqn}) {\em and} (\ref{Ieqn}) fulfills non-trivial Noether
identities.  Note also that in the gauge $g=1$, which is admissible,
the action (\ref{YMaction}) reduces to the massive Yang-Mills action;
the field $g$ appears as a non-abelian St\"{u}ckelberg field.

The other choice is obtained by dropping the Yang-Mills kinetic term
from (\ref{YMaction}), leading to the topological $G/G$-model with
action \begin{equation} {\cal L}=-\frac{1}{2}g_{ab}I^a_{\mu}I^{b\mu}.
\label{topological} \end{equation} The equations of motion are
\begin{eqnarray} I^{a\mu} &=& 0, \label{topo_eqn_of_motion}\\
D^{(\Theta)}_{\mu}I^{a\mu} &=& 0.  \label{topo_eqn_of_motion'}
\end{eqnarray} Again, the $g$-equation of motion is a consequence of
the $A$-equation of motion.  The model has no local degrees of freedom
since in the gauge $g=1$, the connection $A_\mu$ vanishes.

We shall explicitly discuss below these two cases.  However, our
method also covers more general Lagrangians having the same set of
fields and gauge symmetries.  In fact, the explicit form of the
Lagrangian is only used in section \ref{s_mod_d}.  The results of the
following sections are manifestly independent of the dynamics and rely
solely on the form of the gauge symmetries.  And even the results of
section \ref{s_mod_d} are to a large extent independent of the
Lagrangian.

\section{The problem} \setcounter{equation}{0}

The BRST transformation \cite{BRS,T} that incorporates the gauge
symmetries can be constructed by following the general antifield
procedure \cite{Zinn,BV} described for instance in \cite{hen3}.  To
write the BRST-variations of the variables in a convenient form, it is
useful to redefine appropriately the antifields conjugate to the group
element $g$.

In local coordinates, the BRST transformation of the $h^a$ reads
\begin{equation} sh^a=\Omega^a_{\ b}(h)C^b, \end{equation} as it
follows by replacing the gauge parameters $\epsilon^a$ by the ghosts
$C^a$ in the gauge variation of $h^a$.  This term is generated by
taking the antibracket of $h^a$ with $\int d^n x h^*_a \Omega^a_{\ b}
C^b$, which must thus be added to the Yang-Mills solution of the
master equation.  Here, the $h^*_a$ are the antifields conjugate to
$h^a$, \begin{eqnarray} (h^a(x),h^*_b(y))=\delta^a_b\delta(x-y), \\
(h^*_a(x),h^*_b(y))=0.  \end{eqnarray} This implies that the BRST
variation of the antifields $h^*_a$ are given by \begin{equation}
sh^*_a=h^*_b\frac{\delta\Omega^b_{\ c}}{\delta h^a}C^c
+\mbox{equations-of-motion-terms}.  \end{equation} It is possible to
replace the $h^*_a$ by new variables $g^*_a$, with the same gradings,
defined through \begin{equation} g^*_a=h^*_b\Omega^b_{\ a}(h),
\end{equation} which have much simpler BRST transformation rules,
\begin{equation} sg^*_a=g^*_bf^b_{\
ac}C^c+\mbox{equations-of-motion-terms}.  \end{equation} This equation
indicates that the $g^*_a$ transform according to the co-adjoint
representation of ${\cal G}$.  We shall work in the sequel with the
antifields $g^*_a$ rather than $h^*_a$, although they do not have
canonical antibrackets, \begin{eqnarray} (g^*_a,g^*_b) &=& -g^*_c
f^c_{\; ab} \\ (g, g^*_a) &=& -gT_a.  \end{eqnarray} Adopting the
geometrical interpretation of the antifields given in \cite{EdWit},
the $h^*_a$ may be regarded as the vector fields tangent to the
$h^a$-coordinate lines.  Accordingly, they are defined only in the
coordinate patch covered by the $h^a$.  By contrast, the $g^*_a$ are
the left-invariant vector fields and are defined over the entire group
manifold.  In terms of $g^*_a$, the extra term in the solution of the
master equation reads simply $\int d^n x g^*_aC^a$.

We can now write the BRST transformation of all the variables.  Since
the gauge transformations close off-shell, the BRST differential
splits according to the antighost degree in the Koszul-Tate
differential ($\delta$) and the longitudinal differential along the
gauge orbits ($\gamma$):  $s=\delta+\gamma$, with no extra terms.  The
(left) action of these differentials on the fields explicitly reads
\begin{equation} \begin{array}[b]{lllllll} \gamma A^a_{\mu} &=&
\partial_{\mu}C^a+f^a_{\ bc}A^b_{\mu}C^c = D^{(A)}_\mu C^a, &
\mbox{\hspace{0.8cm}} & \delta A^a_{\mu} &=& 0,\\ \gamma g & = & g T_a
C^a, & & \delta g &=& 0,\\ \gamma y^i & = & (Y_a)^i_j y^j C^a, & &
\delta y^i &=& 0,\\ \gamma C^a & = & -\frac{1}{2}f^a_{\ bc}C^b C^c, &
& \delta C^a &=& 0,\\ \gamma A_a^{*\mu} &=& A_b^{*\mu}f^b_{\ ac}C^c, &
& \delta A_a^{*\mu} &=& \frac{\delta S_0}{\delta A^a_{\mu}}, \\ \gamma
g^*_a &=& g^*_bf^b_{\ ac}C^c, & & \delta g^*_a &=& \frac{\delta
S_0}{\delta h^b}\Omega^b_{\ a}(h), \\ \gamma y^*_i &=&
y^*_j(Y_a)^j_iC^a, & & \delta y^*_i &=& \frac{\delta S_0}{\delta y^i},
\\ \gamma C^*_a &=& C^*_b f^b_{\ ac} C^c, & & & & \\ \delta C^*_a &=&
-D^{(A)}_{\mu}A^{*\mu}_a+g^*_a+y_i^*(Y_a)^i_j y^j.  & & & &
\end{array} \label{trafos} \end{equation} These relations imply
\begin{equation} \gamma \Theta_{\mu}^a = D^{(\Theta)}_{\mu} C^a,
\end{equation} and enable us to express the ghost as $C=g^{-1}\gamma
g$.  In the usual abbreviations $C=C^aT_a$, one may rewrite the ghost
transformation law as $\gamma C=-C^2$ since $C^2=\frac{1}{2} T_a
f^a_{\ bc} C^b C^c$.

Our goal is to compute the cohomological groups $H(s|d)$ of the BRST
differential $s$ modulo the spacetime exterior differential $d$, in
the space of local forms.  In ghost degree zero, these groups
characterize the counterterms, while in ghost degree one, they
classify the anomalies.  In negative ghost number, they are related to
the non-trivial conservation laws \cite{hen4}.

The longitudinal derivative $\gamma$ is nilpotent off-shell.
Therefore, we can proceed as in \cite{hen1} and analyse first the
$\gamma$-cohomology, $H(\gamma)$, and the $\gamma$-cohomology modulo
the exterior derivative $d$, $H(\gamma|d)$, in the space of all fields
and antifields.  The De Rham cohomology of the group manifold will
play an important role in this context.  We shall then turn to
$H(s|d)$.

\section{Analysis of $H(\gamma)$} \label{gamma}
\setcounter{equation}{0}

The calculation of the cohomology is performed in the so-called
``jet-space".  This space is simply the (infinite-dimensional) space
coordinatized by the field and antifield components, as well as all
their subsequent partial derivatives, ${\cal K}=\{A^a_{\mu}, g, y^i,
C^a\}$, ${\cal K}^*=\{A_a^{*\mu}, g_a^*, y^*_i, C^*_a\}$,
$\partial_{\mu}{\cal K}$, $\partial_{\mu}{\cal K}^*$ etc.  Because the
spacetime manifold is topologically $R^n$, these functions are
actually globally defined (but note that they do not provide standard
coordinates since the $g$'s are not independent).  The differential
$\gamma$ anticommutes with the exterior derivative, so that the above
transformation laws in Eq.(\ref{trafos}) can be extended to the whole
jet-space.

To describe the $\gamma$-cohomology, it is convenient to employ
different jet-space coordinates.  The construction of these new
coordinates goes as follows.  The quantity $I^a_{\mu}$ defined in
(\ref{I}) can be used instead of the first derivatives of the field
$g$.  Indeed, $I_{\mu}$ can be expressed in terms of $\partial_{\mu}$
and conversly, $\partial_{\mu}g = g(I_{\mu}+A_{\mu})$.  Therefore, the
jet-coordinates $\{g, \partial_{\mu}g, A_{\mu}^a\}$ may be reexpressed
in terms of $\{g, I^a_{\mu}, A_{\mu}^a\}$.  Trying to rearrange the
jet-coordinates with two indices, one finds for the second derivatives
of the group element:  $\partial_{\mu}\partial_{\nu}g = \frac{1}{2}
g(D^{(\Theta)}_{(\mu}I_{\nu)}^aT_a + \partial_{(\mu}A_{\nu)}^aT_a +
\Theta_{(\mu}^a\Theta_{\nu)}^b \{T_a,T_b\} - \Theta_{(\mu}^a
I_{\nu)}^b [T_a,T_b])$.  The derivative of the connection can be split
in a symmetric and an antisymmetric part, $\partial_{\mu}A_{\nu} =
\frac{1}{2}(\partial_{(\mu}A_{\nu)} + \partial_{[\mu}A_{\nu]})$.  But
the curvature ${F}_{\mu\nu} = \partial_{[\mu}A_{\nu]} +
[A_{\nu},A_{\nu}]$ is already contained in the antisymmetrized
$\Theta$-covariant derivatives of $I_{\mu}$,
$D^{(\Theta)}_{[\mu}I^a_{\nu]} = -{F}_{\mu\nu} + [I_{\mu},I_{\nu}]$.
Therefore, only the symmetrized derivatives of $A_{\mu}$ have to be
kept.  {F}urthermore, there are no relations between the new variables
that could constrain the $D^{(\Theta)}_{\mu}I^a_{\nu}$.  Thus, the
coordinates with up to two indices, $\{g, \partial_{\mu}g,
\partial_{\mu}\partial_{\nu}g, A_{\mu}^a, \partial_{\mu}A_{\nu}^a\}$,
may be rearranged to the set $\{g, I^a_{\mu},
D^{(\Theta)}_{\mu}I^a_{\nu}, A_{\mu}^a, \partial_{(\mu}A_{\nu)}^a\}$.

The claim is now that $g$, $A_{\mu}^a$ and all their derivatives can
be replaced by $g$, $A_{\mu}^a$ with its symmetrized derivatives, and
$I^a_{\mu}$ with its successive $\Theta$-covariant derivatives
($k=1,2,\cdots$):  \begin{equation} \Big\{g,\partial_{\alpha_1 \cdots
\alpha_k}g, \partial_{\alpha_1 \cdots \alpha_{k-1}} A_{\alpha_k}^a
\Big\} \rightarrow \Big\{ g, \partial_{(\alpha_1 \cdots \alpha_{k-1}}
A_{\alpha_k)}^a, D^{(\Theta)}_{\alpha_1 \cdots
\alpha_{k-1}}I^a_{\alpha_k} \Big\}.  \end{equation} A good way of
checking the equivalence of the two sets of coordinates is to compare
their size.  Indeed, remembering that the index ``$a$" takes $N$
values and that there are only $N$ independent $g$'s, it is easy to
see that each set contains $N + N\sum^k_{l=1}n(n+1)\cdots (n+l)/l!  +
Nn\sum^{k-1}_{l=1}n(n+1)\cdots (n+l)/l!$ independent coordinates, as
it should be the case ($n$ is the spacetime dimension).  The explicit
proof that the two sets of coordinates are equivalent may be obtained
by induction.  Assume the above statement to be true up to derivatives
of order $k$ for $g$ and of order $k-1$ for $A^a_{\mu}$ (i.e.  for
coordinates with $k$ spacetime indices).  The derivatives of order $k$
of $A^a_{\mu}$ can be expressed in terms of symmetrized derivatives of
$A^a_{\mu}$ and derivatives of order $k-1$ of ${F}_{\mu\nu}^a$.  But
terms of the form $\partial_{\alpha_1 \cdots \alpha_{k-1}}
{F}_{\alpha_k \alpha_{k+1}}^a$ are contained in
$D^{(\Theta)}_{\alpha_1 \cdots \alpha_{k-1}} D^{(\Theta)}_{[\alpha_k}
I^a_{\alpha_{k+1}]}$.  The derivatives of order $k+1$ for $g$ are
generated by taking $k$ symmetrized $\Theta$-covariant derivatives of
$I_{\mu}$:  $D^{(\Theta)}_{(\alpha_1 \cdots \alpha_k}I_{\alpha_{k+1})}
\sim \partial_{(\alpha_1 \cdots \alpha_k}\Theta_{\alpha_{k+1})} +
\mbox{``lower order"} \ \ \sim g^{-1}\partial_{(\alpha_1 \cdots
\alpha_{k+1})}g + \mbox{``l.o."}$, which completes the proof that the
above change of coordinates is indeed invertible.

As new basis of jet-coordinates, we can thus choose the following
combinations of fields and derivatives:  \begin{itemize} \item the
group element $g$ and the ghost $C$ without derivatives, \item the
$I^a_{\mu}$ with all subsequent $\Theta$-covariant derivatives, \item
the matter fields $y^i$ with $\Theta$-covariant derivatives, \item the
antifields ${\cal K}^*$ with $\Theta$-covariant derivatives, \item the
Yang-Mills connection $A^a_{\mu}$ and its symmetrized derivatives, and
the derivatives of the ghost $C$.  \end{itemize} The vector potential
$A^a_{\mu}$ with its symmetrized derivatives and the derivatives of
$C^a$ form contractible pairs, as observed in \cite{DV}.  Accordingly,
they do not contribute to the $\gamma$-cohomology.

The fields $\chi^A := \{I^a_{\mu}, y^i, {\cal K}^*\}$ all transform
linearly under the action of $\gamma$, $\gamma \chi^A \sim (Z_a)^A_B
\chi^B C^a$ (see Eq.(\ref{trafos})).  The $(Z_a)^A_B$ are the
generators of some representation of $G$, for instance to the adjoint
representation in the case of $I^a_{\mu}$.  It is possible to combine
these fields with the group element $g$ to form invariant quantities
$\tilde{\chi}^A = U(g)^A_B \chi^B$, $\gamma\tilde{\chi}^A=0$.  Here,
$U(g)$ stands for the representative of the group element $g$ in the
relevant representation (generated by $Z_a$).  Since $U(g)$ transforms
contragrediently to the corresponding fields or antifields,
\begin{equation} \gamma U(g)=-(-)^{\epsilon_{\chi}}U(g)Z_aC^a,
\label{tilde_variables} \end{equation} the variables $\tilde{\chi}^A$
are invariant, $\gamma \tilde{\chi}^A = 0$, i.e.  one may replace
covariant fields by invariant fields (the $\epsilon_{\chi}$ denote the
parity of the field $\chi$).  {F}urthermore, a short calculation shows
that $\partial_{\mu}\tilde{\chi}^A = U(g)D^{(\Theta)}_{\mu}\chi^A$.
It is therefore possible to replace the jet-variables $\chi^A$ and
their $\Theta$-covariant derivatives by the quantities
$\tilde{\chi}^A$ and their ordinary derivatives.  The introduction of
the tilde variables follows the pattern of \cite{tilde} (see also
\cite{wein2}).

In the new basis of jet-coordinates and after elimination of the
trivial pairs, the action of the longitudinal derivative $\gamma$
reduces to the simple form \begin{eqnarray} \gamma g &=& g C,
\label{trafo1} \\ \gamma C &=& -C^2, \\ \gamma [\tilde{\chi}^A] &=& 0,
\label{trafo3} \end{eqnarray} which fits with the general conditions
on ``good" jet-coordinates given in \cite{BrCMP}.  The square brackets
around $\tilde{\chi}^A$ stand for $\tilde{\chi}^A$ and all the
subsequent ordinary derivatives.  It follows from (\ref{trafo3}) that
the most general solution of the cocycle condition $\gamma m = 0$ is,
up to trivial terms, a linear combination of polynomials in the
gauge-invariant variables $[\tilde{\chi}^A]$ times a solution of
$\gamma n= 0$ involving only the $g$'s and the $C$'s.  To complete the
analysis of the cohomology of $\gamma$, we thus need to compute the
cohomology defined by \begin{eqnarray} \gamma g &=& g C,
\label{DeRhamTransf1} \\ \gamma C &=& -C^2.  \label{DeRhamTransf2}
\end{eqnarray} This is done by relating (\ref{DeRhamTransf1}) and
(\ref{DeRhamTransf2}) to the De Rham cohomology of the group manifold.

It is the identification $\gamma\rightarrow d$ and
$C\rightarrow\Theta$ that establishes the link.  Here, the exterior
derivative $d$ acts in the space of $g$ and $\Theta$ in the same way
as $\gamma$ acts in the space of $g$ and $C$.  Thus the BRST complex
involving the group element and the ghost is identified with the De
Rham complex of the group manifold.  The relevant identities are now
$d g = g \Theta$ and $d \Theta = - \Theta^2$, where the second
equation is recognized to be the Maurer-Cartan structure equation for
left-invariant forms on the group, which we used already above.  Let
$\omega_I=\omega_I(\Theta,g)$ form a basis of $H_{DR}(G)$, and let
$\omega_I(C,g)$ be the function of $C$ and $g$ obtained after
replacing $\Theta$ by $C$ in $\omega_I(\Theta,g)$.  Then, a general
cocycle solving the equations $\gamma m = 0$, has the form
\begin{equation} m = \sum_I P^I([\tilde{\chi}^A],dx) \omega_I(C,g) +
\gamma n, \label{cocycle} \end{equation} where the $P^I$ are arbitrary
polynomials in the variables $[\tilde{\chi}^A]$ and the differentials
$dx^{\mu}$ (we assume no explicit $x$-dependence).  {F}urthermore, $m$
is trivial if and only if $P^I = 0$ (for each $I$).

Note that the invariant polynomials in the covariantly transforming
quantities $\chi^A$, which are related to the Casimir invariants of
the corresponding representation, form a subset of all
$P^I([\tilde{\chi}^A],dx)$.

\section{Topological terms} \label{winding} \setcounter{equation}{0}

Consider the pull-backs to the spacetime manifold of the forms
$\omega_I(\Theta,g)$.  These are just given by $\omega_I(\Theta,g)$
where $\Theta$ is viewed as the spacetime form $\Theta_{\mu}dx^{\mu}$
rather than a $1$-form on the group manifold (and $d$ is the spacetime
$d$ rather than the exterior derivative on the group manifold).  {F}or
this reason, we shall denote these pull-backs by the same symbol
$\omega_I(\Theta,g)$.  The spacetime exterior forms
$\omega_I(\Theta,g)$ are related to the $\gamma$-cocycles
$\omega_I(C,g)$ through the descent equation \cite{hen4}.

Indeed, expanding $\tilde{\omega}_I \equiv \omega_I(\Theta + C,g)$
according to the ghost number yields \begin{equation} \tilde{\omega}_I
= \omega_I^{0,p} + \omega_I^{1,p-1} + \cdots + \omega_I^{p,0}
\end{equation} where $p$ is the form degree of $\omega_I(\Theta,g)$
and where in $\omega_I^{k,l}$, the first superscript $l$ stands for
the form degree while the second superscript $k$ stands for the ghost
number ($k+l=p$).  Of course, $\omega_I^{0,g} = \omega_I(\Theta,g)$
and $\omega_I^{p,0} = \omega_I(C,g)$.  Now, $\tilde{\omega}_I$ is
annihilated by $\tilde{\gamma} = \gamma + d$ by construction,
\begin{equation} \tilde{\gamma} \tilde{\omega}_I = 0 \end{equation}
(the previous equation is usually referred to as ``Russian formula"
\cite{zumi,stora}).  If one also expands this equation according to
the ghost number, one finds a tower of ``descent equations" that read
explicitly \begin{eqnarray} d\omega^{(0,p)} &=& 0 \nonumber\\
\gamma\omega^{(0,p)}+d\omega^{(1,p-1)} &=& 0 \nonumber\\
\gamma\omega^{(1,p-1)}+d\omega^{(2,p-2)} &=& 0 \nonumber\\ &\vdots&
\nonumber\\ \gamma\omega^{(p-1,1)}+d\omega^{(p,0)} &=& 0 \nonumber \\
\gamma\omega^{(p,0)} &=& 0.  \label{descent} \end{eqnarray}

It follows from the Poincar\'e lemma on the group manifold that
$\omega_I^{0,p}$ is locally exact, $\omega_I^{0,p} = d K^{0,p-1}$.
This implies that all the forms $\omega^{k, g-k}$ occuring in the
descent are also locally trivial, $\omega^{k,p-k} = d K^{k,p-k-1} +
\gamma K^{k-1,p-k}$, where $K^{l,p-l}$ is the component of
$\tilde{K}(\Theta +C,g)$ of ghost number $l$.  These relations,
however, hold only locally.  Globally, it is not possible to bring the
$\omega$'s to the trivial form.  {F}or this reason, the $\omega^{k,
g-k}$ will be referred to in the sequel as the ``topological terms".

The descent equations (\ref{descent}) will be exploited in the next
section.  A particularly important case arises when $p=n$.  In that
case, one sees from (\ref{descent}) that the spacetime integral
\begin{equation} \int_{R^n} \omega_I^{0,n} \label{WN} \end{equation}
is gauge-invariant since its integrand is gauge-invariant up to a
total derivative.  It can thus be added to the action without breaking
gauge invariance.  However, because $\omega_I^{0,n}$ is locally exact,
the topological term (\ref{WN}) does not modify the equations of
motion.  The terms of the form (\ref{WN}) are called ``winding number
terms".  Although locally trivial, they cannot be eliminated globally.
Also, their integrands do not differ from a strictly gauge-invariant
integrand up to the exterior derivative of a (globally defined)
($n-1$)-form, since this would imply that the last element in the
corresponding descent is trivial.

{F}or instance in three spacetime dimensions and for a compact, simple
gauge group such as $SU(3)$, the non-trivial $\gamma$-cocycle \[Tr
(g^{-1}\gamma gg^{-1}\gamma gg^{-1}\gamma g)=TrC^3\] corresponds to
the three-form winding number term
\[Tr(g^{-1}dgg^{-1}dgg^{-1}dg)=Tr\Theta^3.\] Varying the field $g$ in
this expression yields a total derivative, which indicates that the
winding number terms do not contribute to the equations of motion for
$g$.  On the other hand, they cannot be globally written as a total
derivative in the space of fields and accordingly, they cannot be
dropped from the action.  Locally, it is of course always possible to
express them as total derivatives.

{F}inally, we note that the forms $\omega_I^{p,0}$ are the only
non-trivial cocycles of the exterior derivative $d$ acting in the
algebra ${\cal A}$ of local forms on the jet-space of the fields,
ghosts and antifields as described in section \ref{gamma}.  This
follows from the generalization of the so-called ``Algebraic
Poincar\'e Lemma" to the case where some fields (here $g$) belong to a
cohomologically non-trivial manifold (here the group manifold $G$)
\cite{vinog,Ander} (see also \cite{hen4}).  \vskip .2cm \noindent {\bf
Algebraic Poincar\'e Lemma.}  The cohomology $H^p(d,{\cal A})$ of $d$
in the algebra of local $p$-forms is isomorphic to the De Rham
cohomology of $G$ in the same form degree for $p<n$, \begin{equation}
H^p(d,{\cal A}) \simeq H^p_{DR}(G), \ \ \ \ p<n.  \end{equation} In
maximal form degree, $H^n_{DR}(G)$ is isomorphic to the quotient of
the variationally closed $n$-forms by the $d$-exact $n$-forms.  An
$n$-form ${\cal L}d^nx$ is said to be variationally closed if and only
if the Euler-Lagrange derivatives of ${\cal L}$ with respect to all
the fields, ghosts and antifields vanish.  \vskip .2cm \noindent {F}or
later purposes, we also quote the Covariant Poincar\'e Lemma, which
describes $H(d)_{inv}$, i.e.  the cohomology of $d$ in the space of
invariant polynomials.  \vskip .2cm \noindent {\bf Covariant
Poincar\'e Lemma.}  Let $P^k([\tilde{\chi}])$ be a $d$-closed
invariant polynomial of form degree $k$.  Then, $P$ may be assumed to
be $d$-exact in the space of invariant polynomials, i.e.
\begin{equation} dP^k([\tilde{\chi}])=0, \ \gamma P=0 \
\Longrightarrow \ P^k = dQ^{(k-1)}([\tilde{\chi}])+\alpha^k, \ \gamma
Q=0, \label{CPL} \end{equation} where $\alpha^k$ is a constant form.
\vskip .2cm \noindent Thus, $H(d)_{inv}$ vanishes in the setting
considered here, contrary to the case without the non-linearly
transforming field $g$, where the obstructions to choosing $Q$
invariant in Eq.(\ref{CPL}) were identified to be the invariant
polynomials in the curvature form ${F}$ \cite{drbr,DV}.  In the
presence of the group valued field $g$, it is however possible to
replace covariant quantities ($\chi$) by invariant ones
($\tilde{\chi}$), and covariant derivatives by ordinary derivatives
(see section \ref{gamma}).  Thus, any polynomial in $[\tilde{\chi}]$
is automatically invariant, and the action of $d$ obviously does not
introduce any new variables.  The vanishing of $H(d)_{inv}$ then
follows because the invariants $\tilde{\chi}^A$ and all their
derivatives are independent jet-variables (subject to no identity).
The effect of the group-valued field is particularly striking in the
Abelian case, where the curvature itself becomes $d$-trivial in the
space of invariants, ${F}=dI$.  Here, $I$ is the gauge invariant
quantity $d\phi - A$, and $g=\exp\{\phi\}$.

\section{Analysis of $H(\gamma|\diff)$} \label{gamma_mod_d}
\setcounter{equation}{0}

The next step towards a complete description of the BRST cohomology
modulo $d$ is the calculation of $H^{(*,*)}(\gamma|d)$.  The
bi-grading ``$(*,*)$" refers as before to the ghost degree and the
form degree respectively.  Via standard descent equations one can,
again as before, relate $H(\gamma|d)$ to the cohomology $H(\gamma)$
which is known from the above analysis.

A representative $a^{(g,p)}$ of some class in $H^{(g,p)}(\gamma|d)$
has to fulfill \begin{equation} \gamma a^{(g,p)} + d a^{(g+1,p-1)} =
0.  \end{equation} If $a^{(g+1,p-1)}$ happens to be trivial in
$H(\gamma|d)$, then it can be eliminated through trivial redefinitions
and $a^{(g,p)}\in H^g(\gamma)$.  If $a^{(g+1,p-1)}$ is not in the
trivial class of $H(\gamma|d)$, then it cannot be trivially absorbed.
In this case, the Algebraic Poincar\'{e} Lemma insures the existence
of a descent \begin{equation} \gamma a^{(g+k,p-k)} + d
a^{(g+k+1,p-k-1)} = 0, \ \ k=1,2,\ldots , \label{generaldescent}
\end{equation} that ends when $a^{(g+k+1,p-k-1)} = 0$ for some value
of $k$, which happens at the latest when zero-forms are produced
\footnote{ {F}or more details on this procedure when the cohomology of
$d$ is non-trivial, as here, see \cite{hen4}.  }.  Any bottom
$a^{g+k,p-k}$ of a descent is a cocycle of $\gamma$, $\gamma
a^{g+k,p-k} = 0$.  Therefore, the last term in the descent takes the
form Eq.(\ref{cocycle}), \begin{equation} a^{g+k,p-k}= \sum_I
P^I([\tilde{\chi}^A],dx) \omega_I(C,g) + \gamma n.  \label{cocycle'}
\end{equation} The $\gamma$-trivial part can be absorbed through
redefinitions of the previous terms and may be assumed to be absent.

It turns out that some non-trivial $\gamma$-cocycles are actually
trivial in $H(\gamma|d)$ and accordingly must also be discarded.  More
precisely, if $P^I$ is $d$-trivial, \begin{equation}
P^I([\tilde{\chi}^A],dx) = d \rho^I([\tilde{\chi}^A],dx), \ \ \gamma
\rho^I = 0, \end{equation} for some invariant polynomial $\rho^I$,
then the corresponding cocycle in $H(\gamma)$ is $\gamma$-trivial
modulo $d$, \begin{equation} a^{(g+k,p-k)} =
P^I([\tilde{\chi}^A],dx)\omega_I(C,g) = d(\rho^I \omega_I) - \gamma
(\rho^I \hat{\omega}_I), \end{equation} where $\hat{\omega}_I$ is the
second to last term in the descent Eq.(\ref{descent}) associated with
the De Rham cohomology of $G$ analysed in the previous section,
$d\omega_I=\gamma\hat{\omega}_I$.

If the descent is non-trivial, so that $a^{g+k,p-k}$ can be lifted at
least once, then $P^I$ must be constant up to terms that are $d$-exact
in the space of invariants.  Indeed, one finds from $d a^{g+k,p-k} +
\gamma a^{g+k-1,p-k+1} = 0$ that $(dP^I) \omega_I + \gamma \mu' =0$,
which yields $dP^I = 0$ since the $\omega_I$ are independent in
cohomology.  The equality $P^I= \alpha^I + d \rho^I$, where the
$\alpha^I$ are constant forms, then follows from the Covariant
Poincar\'{e} Lemma.  As we have just seen, the $d\rho^I$ component of
$P^I$ can be discarded.

The only elements of $H(\gamma)$ that could serve as bottom of a
non-trivial descent are therefore the basis elements of the De Rham
cohomology, $\omega_I$, multiplied by constant $p$-forms.  These
constant forms may be eliminated from the analysis by imposing Lorentz
invariance, which leaves only the zero-forms $\omega_I$ as interesting
bottoms.  {F}urthermore, there is no obstruction to lifting $\omega_I$
up to maximal form degree, as follows immediately from
Eq.(\ref{descent}).  Therefore, any bottom $\alpha^I\omega_I$ is
admissible.

One can summarize the results as follows.  The solutions $a$ of the
cocycle condition $\gamma a + db = 0$ fall into two classes.  {F}irst,
there are the solutions that lead to no (non-trivial) descent, i.e.,
that are strictly annihilated by $\gamma$ (no $d$-exact term occurs),
\begin{equation} \gamma a = 0.  \end{equation} These solutions can be
expressed in terms of the invariants $[\tilde{\chi}^A]$ and the De
Rham forms $\omega_I(C,g)$ as in Eq.(\ref{cocycle'}).  Second, there
are the solutions leading to a non-trivial descent.  These are the
lifts of the De Rham forms $\omega_I(C,g)$, up to trivial terms and
terms strictly annihilated by $\gamma$.  These solutions are locally
trivial (locally in field space) but not globally so.  There are no
other solutions associated with non-trivial descents besides these
topological terms.

\section{Comparison with Pure Yang-Mills Case}
\setcounter{equation}{0}

It is interesting to compare the results obtained here with those of
the pure Yang-Mills case analysed in \cite{DV85,drbr,DV,hen1} (or the
case of Yang-Mills coupled to matter fields transforming according to
some linear representation of $G$).  Since the analysis in those works
was carried out for reductive algebras, we shall assume throughout
this section that ${\cal G}$ is reductive.

In the linear case (by which we mean ``only linear representations"),
the $\gamma$-cohomology is represented by products of elements of the
Lie algebra cohomology with invariant polynomials in the curvature,
the matter-fields, the antifields and the corresponding covariant
derivatives with respect to the gauge connection (denoted by$[ \
]_c$), \[\gamma m=0 \ \Rightarrow \
m=P_{inv}^I([{F}_{\mu\nu}]_c,[y^i]_c,[\Omega^*]_c)\omega^{Lie}_I(C)
+\gamma n.\] Thus, the $\gamma$-cohomologies in the non-linear ($g$
present) and linear ($g$ absent) cases have a similar structure,
except that it is the De Rham cohomology that is relevant in one case,
while it is the Lie algebra cohomology in the other case.  Of course,
for compact groups, the two cohomologies are isomorphic.  But this is
not true in general.

We turn now to the cohomology $H(\gamma|d)$ and assume that the Lie
algebra cohomology and the De Rham cohomology are isomorphic, to
emphasize the differences that arise when working ``modulo $d$".  The
elements of $H(\gamma|d)$ that are not equivalent to elements of
$H(\gamma)$ can be characterized by the bottom of their associated
non-trivial descent, which is a $\gamma$-cocycle.  So, we have to
compare the bottoms that can be lifted in both cases.  We have seen
that in the non-linear case, the only non-trivial bottoms involve only
the ghosts, but no other fields.  This is not true in the linear case,
where one may have bottoms that contain the curvature forms.
Moreover, while there may be obstructions to lifting bottoms more than
once in the linear case \cite{DV85}, this is not true in the
non-linear case, where any bottom can be lifted to maximum form
degree.  {F}or instance, only the primitive elements of the Lie
algebra cohomology can be lifted all the way up to maximum form degree
in the linear case \cite{DV85}.  An example is given by the product
$(Tr C^3)(Tr C^5)$ (in $SU(5)$, say) which cannot be lifted all the
way up to form degree 8 because one encounters the obstruction $Tr
{F}^2$ at form degree 4.  Clearly, $Tr {F}^2$ is non-trivial in
$H(\gamma)$.  When the group-element $g$ is present, then any
combination $\lambda^I\omega_I$ can be lifted up to maximal form
degree.  The product $Tr(C^3)Tr(C^5)$ lifts for example to
$Tr(g^{-1}dg)^3Tr(g^{-1}dg)^5$.  A way to understand the removal of
the obstruction in the non-linear case is to observe that one may lift
the ghosts using the flat connection $\Theta$.  The obstructions are
known to involve the curvatures \cite{DV85}.  They are absent here
because the curvature of $\Theta$ identically vanishes.

{F}inally, we note that in the linear case, the Chern-Simons forms
cannot be replaced by strictly gauge-invariant terms.  By contrast, in
the non-linear case, the Chern-Simons forms differ from strictly
gauge-invariant terms by total derivatives and winding number terms
that are locally trivial and do not contribute to the equations of
motion.  {F}or instance, the Chern-Simons term $3
Tr(AdA+\frac{2}{3}A^3)$ and the winding number term $Tr\Theta^3$
descend on the same cocycle $TrC^3$.  Thus, their difference descends
on zero and hence is equivalent to a strictly gauge invariant term
modulo a total derivative.  An explicit calculation yields indeed
$Tr(AdA+\frac{2}{3}A^3) = Tr\{ID^{\Theta}I-\frac{2}{3}I^3\} -
\frac{1}{3} Tr\Theta^3 -dTr\Theta A$ with $I = I^a_\mu T_a dx^\mu$.

\section{Relation to the work of DeWit, Hull and Ro\v{c}ek}
\label{dewit} \setcounter{equation}{0}

The same conclusions apply to the topological terms considered in
\cite{wit}.  These again differ from strictly gauge invariant terms by
total derivatives (and locally trivial winding number terms if the De
Rham cohomology of the group manifold in form degree $n$ does not
vanish).  {F}rom this point of view, the interesting construction of
\cite{wit} does not bring in new terms in the principal case, even
when the group is not semi-simple.

To illustrate this point, we recall the construction of \cite{wit},
specializing to the principal case and considering four spacetime
dimensions for definiteness.

When trying to construct gauge theories with a non-compact gauge group
$G$, it is natural to consider actions involving integrands of the
form \cite{julia} \begin{equation} {\cal T} = S_{ij}(g) {F}^i \wedge
{F}^j.  \label{cand0} \end{equation} This term is strictly gauge
invariant, $\delta_{\epsilon} (S_{ij}(\phi) {F}^i \wedge {F}^j)=0$, if
$S_{ij}(g)$ transforms in the following way:  \begin{equation}
\delta_{\epsilon} S_{ij} =
-S_{lj}f^l_{ik}\epsilon^k-S_{li}f^l_{jk}\epsilon^k
=-2S_{l(j}f^l_{i)k}\epsilon^k.  \label{sij} \end{equation} When
$S_{ij}$ is an invariant symmetric tensor, the term $S_{ij} {F}^i
\wedge {F}^j$ defines a characteristic class and is a topological
invariant.

In \cite{wit}, De Wit, Hull and Ro\v{c}ek generalize the above setting
through modifications of the action.  As in \cite{vanpr}, they modify
the above term by adding to it an appropriate non gauge-invariant
term, \begin{equation} {\cal T}_{mod}=S_{jk}(g){F}^j\wedge
{F}^k+\frac{2}{3}C_{i,jk}A^i\wedge A^j
(dA^k+\frac{3}{8}f^k_{lm}A^lA^m), \label{cand} \end{equation} and
observe that (\ref{cand}) is invariant up to a total derivative if at
the same time the transformation law for $S_{ij}$, Eq.(\ref{sij}), is
modified to \begin{equation} \delta_{\epsilon}
S_{ij}=(-2S_{l(j}f^l_{i)k}+C_{k,ij})\epsilon^k.  \label{modtrafo}
\end{equation} where the constants $C_{k,ij}=C_{k,ji}$ are subject (i)
to obey $C_{(k,ij)} = 0$ and (ii) to fulfill the $1$-cocycle condition
of the Lie algebra cohomology of ${\cal G}$ in the symmetric tensor
product of the adjoint representation space with istelf \footnote{
{F}or useful information on Lie algebra cohomology, see \cite{azca}.},
$\frac{1}{2}C_{m,ij}f^m_{lk}+f^m_{i[k}C_{l],mj}+f^m_{j[k}C_{l],mi}=0$.
The term involving bare $A$'s in (\ref{cand}) is reminiscent of a
Chern-Simons term.

As the authors of \cite{wit} also observe, any exact contribution to
$C_{k,ij}$ of the form $C_{k,ij}=f^m_{k(i}s_{j)m}$ can be absorbed
through a constant shift of $S_{ij}$.  Thus, if $C_{k,ij}$ is a
coboundary, the term (\ref{cand}) can be brought back to the form
(\ref{cand0}) by redefinition of $S_{ij}$ and addition of a total
derivative and therefore is not a true generalization of
(\ref{cand0}).

Our point is that even when $C_{k,ij}$ is a non-trivial $1$-cocycle of
the Lie algebra cohomology of ${\cal G}$ in the symmetric tensor
product of the adjoint representation space with istelf (which can
only occur when ${\cal G}$ is non semi-simple), one can redefine
(\ref{cand}) (or, for that matter, even (\ref{cand0})) by adding a
total derivative so that this term is strictly gauge invariant and
involves only the manifestly invariant variables $[\tilde{\chi}^A]$
constructed above, up to possible locally trivial winding number
terms.

This is an immediate consequence of our general analysis and is
particularly striking when the gauge group is $R^k$, which is a
non-compact, abelian group.  We denote its generators by $T_a$,
$a=1,\ldots,k$, $[T_a,T_b]=0$.  The field $g$ is then
$\exp[\phi^aT_a]$ where $\phi^a$ is a vector in $R^k$.  The relevant
transformation laws simply read \begin{eqnarray} s\phi^a &=& C^a, \\
sC^a &=& 0, \\ sA^a &=& -dC^a.  \end{eqnarray} As usual, ${F}^a=dA^a$
and $d{F}^a=\delta_{\epsilon}{F}^a=0$.  The De Rham cohomology of $G$
is trivial except for the constants, $H^k_{DR}(G)=0$ for $k\neq0$,
$H^0_{DR}(G)=R$, while the Lie algebra cohomology of ${\cal G}$
consists of the polynomials in the ghosts $C$, $H({\cal G}) = P(C)$.
In particular, there is no winding number term.  {F}urthermore, since
the structure constants are zero, any constant $C_{k,ij}$ with
$C_{k,ij} = C_{k,ji}$ defines a non-trivial $1$-cocycle with value in
the symmetric product of the adjoint representation.  We assume
$C_{(k,ij)}=0$ in the sequel so as to fulfill the first condition (i)
above.

Equations (\ref{cand}) and (\ref{modtrafo}) simplify to
\begin{eqnarray} {\cal
T}_{mod}&=&S_{jk}(\phi){F}^j{F}^k+\frac{2}{3}C_{i,jk}A^iA^j{F}^k,
\label{cand'}\\ sS_{ij}&=&C_{k,ij}C^k.  \end{eqnarray} The
transformation law of $S_{jk}$ implies $S_{jk}(\phi)=C_{i,jk}\phi^i$
up to an irrelevant constant.  The above term ${\cal T}_{mod}$ is
gauge invariant up to a total derivative, $\gamma{\cal
T}_{mod}=-\frac{2}{3}d\{C_{i,jk}(C^iA^j{F}^k+A^iC^j{F}^k)\}$.  {F}rom
the point of view of \cite{wit}, the expression in (\ref{cand'})
represents a non-trivial extension of the strictly gauge invariant
theory, since $C_{k,ij}$ is a non-trivial Lie algebra cocycle.
However, by adding an appropriate total derivative to it, one
straightforwardly verifies that ${\cal T}_{mod}$ is equivalent to the
strongly gauge invariant expression
$\frac{2}{3}C_{i,jk}(d\phi^i-A^i)(d\phi^j-A^j){F}^k$, \begin{equation}
{\cal T}_{mod}\equiv\frac{2}{3}C_{i,jk}\nabla\phi^i\nabla\phi^j{F}^k+
\frac{2}{3}C_{[i,j]k}d(\phi^iA^j{F}^k-\phi^id\phi^j{F}^k),
\end{equation} where $\nabla\phi^i=d\phi^i-A^i$ may be regarded as the
exterior covariant derivative of $\phi$ and is just the invariant
$I^i$ introduced above, $\nabla_\mu \phi^i = I^i_\mu$.

This shows that in the principal case, it is not the Lie algebra
cohomology that controls the ``novelty" of (\ref{cand}).  This term is
always equivalent to a strictly gauge invariant term (plus winding
number terms if $H^4_{DR}(G)$ happens to be non-trivial).  It would be
interesting to extend the analysis of this issue to scalar fields
taking values in quotient spaces $G/H$, for which the general
construction of \cite{wit} was devised.

\section{Gauged Wess-Zumino-Witten term} \setcounter{equation}{0}

The above calculation of $H(\gamma \vert d)$ sheds also a new light on
the problem of gauging the Wess-Zumino-Witten term
\cite{wess,WZW,hull,fig}.  The Wess-Zumino-Witten term ${\cal
L}_{WZW}(g)$ is a term that can be added to the Lagrangian of the
(ungauged) non-linear $\sigma$-model without breaking its rigid
symmetries.  Its characteristic property (which may be used as its
definition) is that it is not strictly invariant under the rigid
symmetries of the model, but only invariant up to a surface term.
{F}urthermore, one cannot ``improve" it by a surface term such that
the sum is strictly invariant (even locally in field space).

Because the Wess-Zumino-Witten term is invariant only up to a
non-trivial surface term, its gauging raises difficulties.  These have
been analysed in \cite{hull,fig}, with the conclusion that in the
principal case in which one gauges the right action (as here), there
are unremovable obstructions to gauging the Wess-Zumino-Witten term.
These obstructions have been related in \cite{fig} to the equivariant
cohomology.  The impossibility of gauging the Wess-Zumino-Witten term
is also a direct consequence of our analysis.

Indeed, suppose that one has found a functional ${\cal
L}_{WZW}([g],[A^a_{\mu}])$ that (i) reduces to the Wess-Zumino-Witten
term ${\cal L}_{WZW}(g)$ when the gauge field is set to zero and (ii)
is gauge-invariant up to a surface term, \begin{equation} \gamma{\cal
L}_{WZW}([g],[A^a_{\mu}]) +da^{(0,n-1)}([g],[A^a_{\mu}],C)=0.
\end{equation} Such a term would provide a ``gauging" of the WZW term.
But our results indicate that such a term would necessarily be
equivalent to a strictly invariant term, modulo winding number terms
that do not contribute to the equations of motion, \begin{equation}
{\cal L}_{WZW}([g],[A^a_{\mu}]) = {\cal L}^{inv}([g],[A^a_{\mu}]) + dm
+ \hbox{ ``winding number terms"} \end{equation} for some $m$.  This
would imply, upon setting $A^a_\mu$ and its subsequent derivatives
equal to zero, that the original Wess-Zumino term is (locally)
equivalent to the strictly invariant term ${\cal
L}^{inv}([g],[A^a_{\mu}]=0)$, which we know cannot be true.  (The
strict invariance of ${\cal L}^{inv}([g],[A^a_{\mu}]=0)$ under rigid
transformations follows from the strict invariance of ${\cal
L}^{inv}([g],[A^a_{\mu}])$ under gauge transformations.)  This means
that there simply is no room for a gauged Wess-Zumino-Witten term.

Our approach is less explicit than the analysis of \cite{fig} since it
does not identify the nature of the obstruction (it just indicates
that there is an obstruction).  At the same time, it is more complete
because we show that the obstruction exists even if one allows ${\cal
L}_{WZW}$ to depend on the individual field components and all their
derivatives.  As pointed out very clearly in \cite{fig}, the previous
calculations were performed only in the ``universal" algebra generated
by $g$, the $1$-form $A$ and their exterior derivatives $dg$, $dA$
(but not in the algebra generated by all the separate individual
components of the fields and their higher order derivatives).  So,
these calculations excluded only gauged Wess-Zumino-Witten terms
${\cal L}_{WZW}$ depending on $g$, $A$ and their exterior derivatives
but still left open the possibility of gauging ${\cal L}_{WZW}$ in the
``big algebra" containing all the field components and their
derivatives individually \cite{fig}.

\section{Analysis of $H(\lodiff|\diff)$} \label{s_mod_d}
\setcounter{equation}{0}

In order to characterize the cohomology modulo $d$ of the complete
BRST operator in the space of fields and antifields, it is necessary
to specify the dynamics of the theory.  Indeed, $s$ contains
information on the equations of motion through the Koszul-Tate
differential $\delta$, and the BRST cohomology will in general depend
on the dynamics although the gauge transformations are not affected.
We shall first develop the analysis in the case of the usual action
(\ref{YMaction}), which is, if one reinstates explicitly the coupling
constants, \begin{eqnarray} {\cal
L}&=&-\frac{1}{4}g_{ab}{F}^a_{\mu\nu}{F}^{b\mu\nu} -m^2
\frac{1}{2}g_{ab}I^a_{\mu}I^{b\mu} + \mbox{matter action}
\label{standardaction}\\ {F}^a_{\mu\nu} &=& \partial_\mu A^a_\nu -
\partial_\nu A^a_\mu + \alpha f^a_{\ bc} A^b_\mu A^c_\nu \\ I^a_{\mu}
&=& (\omega^a_{\,b}(h)\partial_\mu h^b - \alpha A^b_\mu)
\end{eqnarray} where $\alpha$ is the Yang-Mills coupling constant.  We
shall then explain how the results extend to more general actions.

The idea follows the pattern developed in \cite{hen1,hen2}.  One
controls the antifield dependence of the BRST cocycles through
expansion of the condition \begin{equation} s a + d b = 0
\end{equation} according to the antighost number, $a=a_0+\cdots+a_k$
and $b=b_0+\cdots+b_m$.  Only the case where the highest antighost
degree of $a$ is equal to that of $b$ ($k=m$) shall be described here
because the other cases can be easily reduced to this one.  At highest
antighost number $k$ -- which we take to be $>0$ since otherwise there
is no antifield --, the above cocycle condition reads \begin{equation}
\gamma a_k + d b_k = 0.  \end{equation} This implies $\gamma b_k +
dc_k = 0$ and hence, according to our analysis of $H(\gamma|d)$, $b_k$
must be trivial (it must be liftable at least once but it contains the
antifields and therefore cannot be a pure topological term).  We can
thus assume $\gamma a_k =0$, i.e., up to trivial redefinitions,
$a_k=P^I\omega_I$.  Next, the subleading equation in the above
decomposition of the cocycle condition has to be used:
\begin{equation} \delta a_k + \gamma a_{k-1} + d b_{k-1} = 0.
\label{subleading} \end{equation} Acting with $\gamma$ on this
equation produces $d\gamma b_{k-1}=0$ and thus $\gamma b_{k-1}+d
c_{k-1}=0$.  By the same reasoning as above, one finds that $c_{k-1}$
is trivial if $k>1$, and thus one may assume $b_{k-1}=Q^I\omega_I$.
If $k=1$, $b_0=b_0^{inv}+b_0^{top}$ may have a non-trivial,
topological component.  The resulting equation $\gamma
b_0^{top}+dc_0=0$ may be lifted to $\gamma a_0^{top}+db_0^{top}=0$.
By subtracting, if necessary, the topological term $a_0^{top}$ from
$a_0$, it is possible to eliminate the non-invariant component of
$b_0$ and to assume $b_{k-1}=Q^I\omega_I$ also for $k=1$.

Upon inserting the explicit forms of $a_k$ and $b_{k-1}$ in
Eq.(\ref{subleading}), it is straightforward to derive that $\delta
P^I + d Q^I = 0$.  {F}urthermore, if $P^I$ is in the trivial class of
$H^n_k(\delta|d)$, i.e.  if $P^I=\delta M^I+d N^I$, then $a_k$ can be
absorbed through trivial redefinitions.  The antighost dependence of
$a$ is thus controlled by $H^n_k(\delta|d)$.  It is through these
cohomological groups that the dynamics enter.

The group $H^n_k(\delta|d)$ has been shown in \cite{hen5} to be
isomorphic to the characteristic cohomology $H^{n-k}(d|\delta)$ of
antifield-independent $(n-k)$-forms that are weakly closed (i.e.
closed modulo the equations of motion) but not weakly exact.  Thus,
$H^n_1(\delta|d)$ is isomorphic to the space of non-trivial weakly
conserved currents.  It does not vanish for the above theory, which is
Poincar\'e invariant.  {F}or higher antighost degree $k>1$, the groups
$H^n_k(\delta|d)$ turn out, however, to be trivial \cite{hen5}.

The triviality of $H^n_k(\delta|d)$ for $k>2$ follows from the general
theorems of \cite{hen5,vinog,brya}.  The triviality of
$H^n_2(\delta|d)$ is demonstrated by following the perturbative
argument of \cite{hen5}:  the theory obtained by taking the limit
$\alpha = 0$ in the action (\ref{standardaction}) describes a set of
$U(1)$ gauge fields together with a non-linear $g$-field with rigid
$G$-symmetry, which does not interact with the gauge
fields\footnote{The corresponding equations of motion are obtained by
keeping in the original equations of motion the terms with the maximum
number of derivatives.  Thus, the perturbative method of \cite{hen5}
indeed applies.}.  In that limit, the only non-trivial cocycles of
$H^n_2(\delta|d)$ are known to be $\lambda^a C^*_a$ up to trivial
terms.  These terms cannot be deformed to cocycles of the full theory
when $\alpha \not= 0$, even in the abelian case (the undifferentiated
term $g^*_a$ in $\delta C^*_a$ prevents it) and thus we may conclude
that $H^n_2(\delta|d)$ vanishes (we refer the reader to \cite{hen5}
for a detailed explanation of the method).

Note also that if $P^n_k([\tilde{\chi}^A])$ is a trivial invariant
polynomial, $P^n_k([\tilde{\chi}^A])=\delta M^n_{k+1}+ dN^{n-1}_k$,
then it is also trivial in the space of {\em invariant} polynomials,
as one can see by setting all variables equal to zero in $M^n_{k+1}$
and $N^{n-1}_k$ but the gauge invariant $[\tilde{\chi}^A]$.  In
contrast to the situation analysed in \cite{hen1}, the invariance of
$M^n_{k+1}$ and $N^{n-1}_k$ is thus not an issue here.

Let us come back to the analysis of the cocycle condition $sa + db
=0$.  The fact that the only non-vanishing cohomology group
$H^n_k(\delta|d)$ is $H^n_1(\delta|d)$ implies that the BRST cocycles
may be assumed to have an expansion that stops after the second
summand, $a = a_0 + a_1$, where $a_1$ may be chosen invariant,
$a_1=P^I\omega_I$, and $P^I\in H^n_1(\delta|d)$.  If $gh(a)<0$, then
of course $a=a_1$ (and $gh(a)$ is actually equal to $-1$).  There are
thus two types of cocycles in $H(s|d)$:  those for which $a_1$ does
not vanish (they involve non-trivially the antifields), and those for
which $a_1=0$.  We shall call the first class ``type I", while
solutions in the second class will be of ``type II".

The analysis of the BRST cohomology for other gauge-invariant
Lagrangians proceeds in exactly the same fashion.  If these gauge
invariant Lagrangians fulfill the rather mild ``normality condition"
given in \cite{hen5}, the groups $H^n_k(\delta|d)$ are also zero for
$k>1$.  Thus, the solutions of the BRST cocycle condition $sa +db=0$
fall again in two classes, just as for the specific Lagrangian
(\ref{YMaction}).  \vskip .3cm \noindent {\bf Type I}

Let $\{j^\mu_A\}$ be a complete set of gauge-invariant conserved
currents and let $c_A$ be such that \begin{equation} \partial_\mu
j^\mu_A+\delta c_A=0, \ \gamma c_A = 0, \ antigh(c_A)=1 \end{equation}
(the $c_A$'s define the rigid symmetries associated with the conserved
current $j^\mu_A$ \cite{hen5}).  The solutions of type I take the form
\begin{equation} k^A_I ( j^\mu_A \hat{\omega}^I_\mu + c_A \omega^I),
\end{equation} where the $\omega^I(g,C)$ are the De Rham cocycles and
$\gamma \hat{\omega}^I_\mu + \partial_\mu \omega^I = 0$.  In order to
completely list all the independent solutions of type I, it is
necessary to know all the local conserved currents.  This is a
question that depends on the detailed form of the Lagrangian and that
will not be pursued here.  Two remarks should however be made:  (i)
Potential anomalies are classified through $H^{(1,n)}(s|d)$.  The
above results indicate that there is no anomaly of type I, i.e.  that
the antifield dependence of anomalies may be eliminated through
trivial redefinitions if $H^2_{DR}(G) = 0$, no matter what the
conserved currents are.  In a similar manner, one can get rid of the
antifields in $H^{(0,n)}(s|d)$ if $H^1_{DR}(G)=0$.  $H^{(0,n)}(s|d)$
classifies the observables of the theory, and is relevant for
renormalization and deformation issues.  (ii) The solutions of type I
become trivial upon restricting the fields to lie in a neighbourhood
of the identity since the forms $\omega^I$ are then trivial.  Again,
this is true independently of the form of the conserved currents.
\vskip .3cm \noindent {\bf Type II}

The solutions of type II do not involve the antifields.  The BRST
cocycles $sa+db = 0$ are then $\gamma$-cocycles, $\gamma a +db = 0$
($sa = \gamma a$).  As we have seen, the solutions of this latter
equation also fall into two classes:  those that are strictly
invariant and those that are invariant only modulo a total derivative,
the so-called topological terms (see section \ref{winding}).  Although
the cocycle condition of the $s$ mod $d$ cohomology reduces to the
cocycle condition of the $\gamma$ mod $d$ cohomology when $a$ does not
contain the antifields, the coboundary condition is different.  Some
classes of $H(\gamma \vert d)$ are trivial in $H(s \vert d)$, namely,
those that are zero when the equations of motion hold (or more
generally, $\delta$-exact).  The dynamics plays thus a central
r{\^o}le for determining the explicit form of the most general
coboundary of type I.  This is particularly obvious in the topological
$G/G$-model, to which we now turn.

\section{$G/G$ Topological theory} \setcounter{equation}{0}

{F}or the topological action Eq.(\ref{topological})
\cite{topo1,topo2,topo3,topo4,topo5,topo6}, the local BRST cohomology
$H(s|d)$ reduces to the topological terms of section \ref{winding}.
There is no other cohomological class.  The most expedient way to see
this is to redefine the gauge-invariant $[\tilde{\chi}]$ variables
(see Eq.(\ref{tilde_variables})) in such a way that they form
contractible pairs.

With the definition \begin{equation} \tilde{g}'^*_a = \tilde{g}^*_a -
\partial_\mu \tilde{A}^{*\mu}_a, \end{equation} the $s$-variations of
the new tilde variables simply become \begin{eqnarray}
s\tilde{A}^{*\mu}_a = \tilde{I}^\mu_a \,&,& \; \; s\tilde{I}^\mu_a = 0
\\ s\tilde{C}^*_a = \tilde{g}'^*_a \, &,& \; \; s\tilde{g}'^*_a = 0.
\end{eqnarray} {F}or deriving the previous equations, one has to take
into account
Eqns.(\ref{topo_eqn_of_motion}-\ref{topo_eqn_of_motion'}), as well as
the interchangebility of $D^{(A)}_{\mu}$ and $D^{(\Theta)}_{\mu}$
acting on $I^{\mu}$.  Thus, the gauge-invariant variables
$\tilde{\chi}$ and their derivatives all drop out from the BRST
cohomology, leaving only the undifferentiated group element $g$ and
the ghost $C^a$, the BRST transformations of which are
\begin{equation} sg = gC, \ \ sC= -C^2.  \end{equation} Accordingly,
only the cocycles $\omega^I(g,C)$ and their lift appear in the BRST
cohomology $H(s|d)$.  In particular, the only non-trivial local
observables are the winding numbers.

\section{Perturbation theory} \setcounter{equation}{0}

The De Rham cohomology detects the global properties of the group
manifold $G$.  It is customary, in the context of perturbation theory,
to restrict the fields $g$ to a neighbourhood of the identity of $G$
homeomorphic to $R^k$ (denoted by $\tilde{G}$ in the sequel).  Then,
$H^k_{DR}(\tilde{G})=0$ for $k\neq0$ and $H^0_{DR}(\tilde{G})=R$.
This greatly simplifies the analysis.

{F}irst, one finds that the ghosts $C^a$ drop out from the cohomology.
Indeed, one may redefine in $\tilde{G}$ the ghosts as $C^a \rightarrow
D^a= \Omega^a_{\, b}(h) C^b$.  These new variables form contractible
pairs with the $h^a$, \begin{equation} \gamma h^a = D^a, \; \gamma D^a
= 0 \end{equation} and also \begin{equation} s h^a = D^a, \; s D^a =
0.  \end{equation} Thus, $H(\gamma)$ is given by the functions of the
$\tilde{\chi}$ and their derivatives as observed in \cite{Drago}.
This implies that $H^k(\gamma)$ vanishes unless $k<1$.  The De Rham
cocycles $\omega^I(g,C)$ are trivial in $\tilde{G}$.

Second, because the ghosts drop out from $H(\gamma)$, only the
cocycles of one type survive in $H(\gamma|d)$, namely those that lead
to trivial descents and that can be redefined so as to be strictly
annihilated by $\gamma$.  The topological cocycles disappear.  At
ghost number zero, the terms that are gauge invariant only up to a
total derivative can thus be replaced by strictly gauge invariant
terms involving only the $\tilde{\chi}$-variables and their
derivatives.

{F}inally, only the cohomological groups $H^{(0,n)}(s|d)$ and
$H^{(-1,n)}(s|d)$ are different from zero.  This is again because the
ghosts drop out from the cohomology.  Hence, in the expansion of the
BRST cocycles $a$ ($sa+db=0$) according to the antighost number, one
may assume that there is only one term, $a= a_k$, with $gh(a) = -k= -
antigh(a)$, $\gamma a_k =0$, $\delta a_k + db_{k-1} = 0$.  Non-trivial
solutions are obtained only for $k=0,1$.  The solutions with $k=1$
correspond to the gauge invariant conserved currents considered above.
The solutions with $k=0$ are the observables and can be assumed to be
strictly gauge invariant, i.e.  to involve only the $[\tilde{\chi}]$
(note again that the condition $\delta a_k + db_{k-1} = 0$ is empty
for $k =0$ since $a$ contains then no antifield, but that the
coboundary condition is non-trivial and eliminates the on-shell
vanishing observables).

In particular, there is no perturbative anomaly.  This provides a
cohomological interpretation of the Wess-Zumino anomaly cancellation
mechanism \cite{wess,zumi,frolo}.  By enlarging the original field
space with the group elements $g$ (if the complete gauge group is
broken), the anomaly becomes trivial, i.e.  eliminable through a local
counterterm.  In the antifield language, this means that there exists
a local counterterm $M_1$ which trivializes the anomaly $\Delta S$,
$\gamma M_1=(M_1,S)=i\Delta S$ \cite{brag,gomi,troo}.

\section{Global ghosts} \setcounter{equation}{0}

{F}inally, the situation of a non-gauged sigma model shall be
considered.  The theory contains only the group elements $g\in G$ and
is invariant under the global transformation $\delta_{\epsilon}g =
gT_a\epsilon^a$, where $\epsilon^a$ are constant parameters.  The main
interest of this setting lies in the construction of effective
actions, where it is crucial to have an exhaustive list of all
operators that are compatible with the rigid symmetries (see e.g.
\cite{wein2}).

The incorporation of rigid symmetries in the antifield formalism has
been analysed in \cite{glob1,glob2} in the context of the sigma model.
{F}urther developments may be found in \cite{glob3}.  The symmetry
parameters $\epsilon^a$ are promoted to anticommuting constant ghosts
$C^a$ and the relevant transformation laws read \begin{eqnarray}
\hat{\gamma} g &=& gT_aC^a, \\ \hat{\gamma} C^a &=& -\frac{1}{2}f^a_{\
bc}C^bC^c.  \end{eqnarray} The aim is now to compute the cohomology of
$\hat{\gamma}$ in the set of fields \begin{equation}
\{C^a,g,\partial_{\mu}g,\partial_{\mu}\partial_{\nu}g,\ldots\}.
\end{equation} Derivatives of the global ghosts obviously cannot occur
since they are zero.  As before, all the derivatives of $g$ may be
reexpressed through the variables $\Theta = g^{-1}dg$ and their
subsequent derivatives, yielding as new coordinates of the jet-space
the set \begin{equation}
\{C^a,g,\Theta_{\mu},\partial_{\mu}\Theta_{\nu},\ldots\},
\end{equation} or equivalently, using the invariant tilde variables
$\tilde{\Theta}^a=U(g)^a_b\Theta^b$, \begin{equation}
\{C^a,g,\tilde{\Theta}_{\mu},\partial_{\mu}\tilde{\Theta}_{\nu},\ldots\}.
\end{equation} In these variables, the action of $\hat{\gamma}$ takes
the simple form \begin{eqnarray} \hat{\gamma} g &=& gT_aC^a, \\
\hat{\gamma} C^a &=& -\frac{1}{2}f^a_{\ bc}C^bC^c, \\ \hat{\gamma}
[\tilde{\Theta}^a] &=& 0.  \end{eqnarray} The first two equations can
again be identified with the De Rham complex while the last equation
states that the $[\tilde{\Theta}]$ are invariant.  The representatives
of the $\hat{\gamma}$-cohomology have thus the form found above,
\begin{equation} m=P^I([\tilde{\Theta}])\omega_I(C,g),
\label{globalcocycle} \end{equation} where the $P^I$ are arbitray
polynomials and the $\omega_I$ form a basis of the De Rham cohomology
of $G$.

Apart from the strictly invariant terms, which are exhaustively
classified by Eq.(\ref{globalcocycle}), also the invariant terms that
are invariant only up to a total divergence play an important role in
various physical models.  These terms can be analysed via descent
equations in almost the same way as in section \ref{gamma_mod_d}.  A
non-trivial solution of the mod $d$ cocycle condition at form degree
$n$, $\hat{\gamma}a^{(g,n)}+da^{(g+1,n-1)}=0$, necessarily descends
all the way down to zero-forms as in Eq.(\ref{generaldescent}).  But
now, in the last step of the descent, the constants cannot be
discarded any more.  Indeed, one gets at the last step the condition
$d\hat{\gamma}a^{(g+n,0)}=0$.  It follows from the Algebraic
Poincar\'{e} Lemma that $\hat{\gamma}a^{(g+n,0)}$ has to be equal to a
constant, which must be of ghost degree $g+n+1$.  In the gauged case,
there was no such constant since the ghosts are fields.  Here,
however, the ghosts are constant and so, $\hat{\gamma}a^{(g+n,0)}$ may
be a polynomial in $C^a$.  \begin{equation}
\hat{\gamma}a^{(g+n,0)}(g,C)=\alpha(C).  \label{globaldescent}
\end{equation} This phenomenon was previously observed in a similar
context in \cite{fried}.  By applying $\hat{\gamma}$ to
Eq.(\ref{globaldescent}) it follows that $\hat{\gamma}\alpha(C)=0$.
If in addition $\alpha(C)$ is $\hat{\gamma}$-trivial in the space of
constants, $\alpha(C)=\hat{\gamma}\beta(C)=0$, then it may be absorbed
by trivial redefinitions of the preceding descent equations.  In that
case, the bottom $a^{(g+n,0)}$ fulfills $\hat{\gamma}a^{(g+n,0)}(g,C)=
0$ and thus, as we have seen in section \ref{gamma_mod_d}, is
equivalent to a De-Rham cocycle $\omega_I(\Theta,g)\in
H_{DR}^{g+n}(G)$.  {F}or $g=0$, these cocycles lift up to winding
number terms in form degree $n$.

Upon restricting the $g$-field to a neighbourhood $\tilde{G}$ of the
identity, the De Rham cocycles become trivial and accordingly can be
absorbed through redefinitions.  Thus, if $\alpha(C)$ vanishes in
(\ref{globaldescent}), or is $\hat{\gamma}$-trivial in the space of
constants, the original $a^{(g,n)}$ differs from a term strictly
annihilated by $\gamma$ by a total divergence.  The obstruction to
replacing a term invariant up to a total divergence in the Lagrangian
by a term strictly invariant is thus an element of the Lie algebra
cohomology $H({\cal G})$:  if one hits a non-trivial Lie cocycle
$\alpha(C)$ in the descent, there is no way to redefine the Lagrangian
so that it is strictly invariant.

{F}urthermore, any Lie algebra cocycle can be written, in the
neighbourhood of the identity, as $\hat{\gamma}a^{(g+n,0)}(g,C)$ for
some $a^{(g+n,0)}$ that involves explicitly the $\sigma$-field $g$.
This is because the De Rham cohomology of $\tilde{G}$ is trivial.
Replacing $C$ by $g^{-1}(d+\hat{\gamma})g$ in $a^{(g+n,0)}(g,C)$ and
keeping the term of form degree $n$ yields the top of a descent
generating $\alpha(C)$ at the bottom.  Thus, any Lie algebra cocycle
$\alpha(C)$ can be lifted all the way up to form degree $n$.  (On the
full group manifold $G$, the term $a^{(g+n,0)}(g,C)$ will in general
not be globally defined.  This leads to a multiply-valued Lagrangian
with a quantization condition on the corresponding coupling constant.)
It follows that the local $n$-forms with vanishing ghost degree that
are invariant up to a divergence are classified by the Lie algebra
cohomology at ghost degree $n+1$, $H^{n+1}({\cal G})$.

{F}or compact groups, the Lie algebra cohomology is isomorphic to the
De Rham cohomology of the (complete) group manifold $G$, which
establishes the link to the results of D'Hoker and Weinberg
\cite{wein3} (see also \cite{AzMacPer}).

\section{Conclusions} \setcounter{equation}{0}

In this paper, we have investigated the local cohomology of the gauged
principal non-linear sigma-model.

The analysis has been pursued by taking due account of the topology of
the group manifold.  We have shown that the most general local BRST
cocycle $a$ is, up to trivial contributions, the sum of terms of three
different kinds, \begin{equation} sa+db=0 \Longleftrightarrow
a=A_1+A_2+A_3+sm+dn.  \end{equation} The cocycle $A_1$ has been called
of ``type I" and involves the antifields linearly, as well as the
conserved currents.  The cocycles $A_2$ and $A_3$ do not involve the
antifields and are of ``type II".  $A_2$ is strictly annihilated by
$\gamma$ and involves therefore only the gauge invariant variables
$\tilde{\chi}^A$ and their derivatives.  The cocycle $A_3$ depends on
$g$ and the ghosts.  It is a solution of $\gamma A_3+db=0$ and is
related to the De Rham cohomology of the group manifold.  $A_1$ is
also related to $H^k_{DR}(G)$, so that both, $A_1$ and $A_3$ may be
regarded as ``generalized winding number terms".

At ghost number 0 and form degree $n$ (observables), $A_3$ exists if
and only if $H^n_{DR}(G)\neq 0$.  Similarly, $A_1$ exists if and only
if $H^1_{DR}(G)\neq 0$ and if there are non-trivial conserved
currents.  $A_3$ defines a term which is gauge invariant up to total
derivatives, $A_1$ defines a term which is gauge invariant up to field
equations and total derivatives, while $A_2$ defines a term which is
strictly gauge invariant.

In perturbation theory, it is customary to replace $G$ by a
topologically trivial neighbourhood $\tilde{G}$ of the identity.  In
doing so, both $A_1$ and $A_3$ disappear at ghost number 0, and only
the strictly gauge invariant terms are left.  {F}urthermore, there is
no cohomology at positive ghost number.  In particular, there is no
non-trivial anomaly.

It would be interesting to extend the analysis to coset models built
on a homogeneous space $G/H$.  Work in this direction is in progress.

\section*{Acknowledgements}

Discussions with Ricardo Amorim, Glenn Barnich, Nelson Braga,
{F}riedemann Brandt, Michel Dubois-Violette, Antonio Garc{\'\i}a,
Bernard Julia and Claude Viallet are gratefully acknowledged.

\end{document}